\documentclass[11pt]{article}

\usepackage{lipsum}    % For generating dummy text
\usepackage{siunitx}  % Include this in your preamble
\usepackage{graphicx}  % Required for including images
\usepackage{float}     % Improved interface for floating objects
\usepackage{comment}
\usepackage{amsmath}  % Essential for advanced equations
\usepackage{hyperref} % for hyperlinks and PDF metadata
\usepackage[labelfont=it]{caption} % italic captions
\usepackage{url} % for handling URLs
\usepackage{xcolor} % for color handling
\usepackage{geometry} % for page dimension and margin setting
\usepackage{authblk} % for author affiliations
\usepackage{multicol} % for multi-column layout
\usepackage[labelfont=it]{caption}
\usepackage{url}
\usepackage{xcolor}
\usepackage{hyperref}
\usepackage{geometry}
\usepackage{float}
\usepackage{amsmath}
\usepackage{amssymb}
\usepackage{graphicx}
\usepackage{mathtools}
\usepackage[nottoc]{tocbibind}
\usepackage{verbatim}
\usepackage[export]{adjustbox}
\usepackage{graphics}
\usepackage{array}
\usepackage{titlesec}
\titlespacing*{\section} {0pt}{1ex}{1ex}    % this is used to reduce the spacing around the section
\titlespacing*{\subsection} {0pt}{1ex}{1ex}    % this is used to reduce the spacing around the section

\expandafter\def\expandafter\normalsize\expandafter{%
    \normalsize%
    \setlength\abovedisplayskip{4pt}%
    \setlength\belowdisplayskip{8pt}%
    \setlength\abovedisplayshortskip{-4pt}%
    \setlength\belowdisplayshortskip{5pt}%
}   % this is to reduce the space around equations
\expandafter\def\expandafter\small\expandafter{%
    \normalsize%
    \setlength\abovedisplayskip{-10pt}%
    \setlength\belowdisplayskip{5pt}%
    \setlength\abovedisplayshortskip{-10pt}%
    \setlength\belowdisplayshortskip{0pt}%
}   % this is to reduce the space around equations

\title{\vspace{-2em}First Ion Temperature Measurements in the MAST-U Divertor via Retarding Field Energy Analyzer}

\author[1,2]{Y. Damizia}
\author[2]{S. Elmore}
\author[2]{P. Ryan}
\author[2]{S. Allan}
\author[3]{F.Federici}
\author[1,2]{N.Osborne}
\author[1]{J. W. Bradley}
\author[2]{the MAST-U Team}

\affil[1]{Electrical Engineering and 
Electronics, University of 
Liverpool, Liverpool, L69 3GJ, UK}

\affil[2]{UK Atomic Energy Authority, Culham Campus, Abingdon, OX14 3DB, UK%\\This line break forced with \textbackslash\textbackslash
}%

\affil[3]{ 
Oak Ridge National Laboratory, Oak Ridge, Tennessee 37831, USA}%\\This line 
\date{\vspace{-5ex}} % Reduce space after the title block

\begin{document}

\maketitle
\begin{abstract}
\noindent % Ensures the abstract is not indented

This study presents the first ion temperature (\(T_i\)) measurements from the MAST-U divertor using a Retarding Field Energy Analyzer (RFEA). Embedded within the flat tile of the closed divertor chamber, the RFEA captures \(T_i\) profiles across various plasma scenarios, including transitions to the Super-X configuration. Measurements were conducted under steady-state and transient plasma conditions characterized by a plasma current (\(I_p\)) of 750 kA, electron density (\(n_e\)) between \(2.2 \times 10^{19}\) and \(4.45 \times 10^{19}\,\text{m}^{-3}\), and Neutral Beam Injection (NBI) power ranging from 3.0 MW to 3.2 MW. The ion temperatures, peaking at approximately 17 eV in steady state, were compared with electron temperatures (\(T_e\)) obtained from Langmuir probes (LP) at identical radial positions. Preliminary findings reveal a \(T_i/T_e\) ratio ranging from 1 to 2.2. Additionally, high temporal resolution measurements (100 $\mu s$) captured the dynamics of Edge Localized Modes (ELMs), showing \(T_i\) peaks at 16.03 ± 1.84 eV during ELM events, nearly three times higher than inter-ELM temperatures.
\end{abstract}

\begin{multicols}{2} % Begin two-column layout

\section{Introduction}
\begin{figure}[H]
\centering
\includegraphics[width=0.48\textwidth]{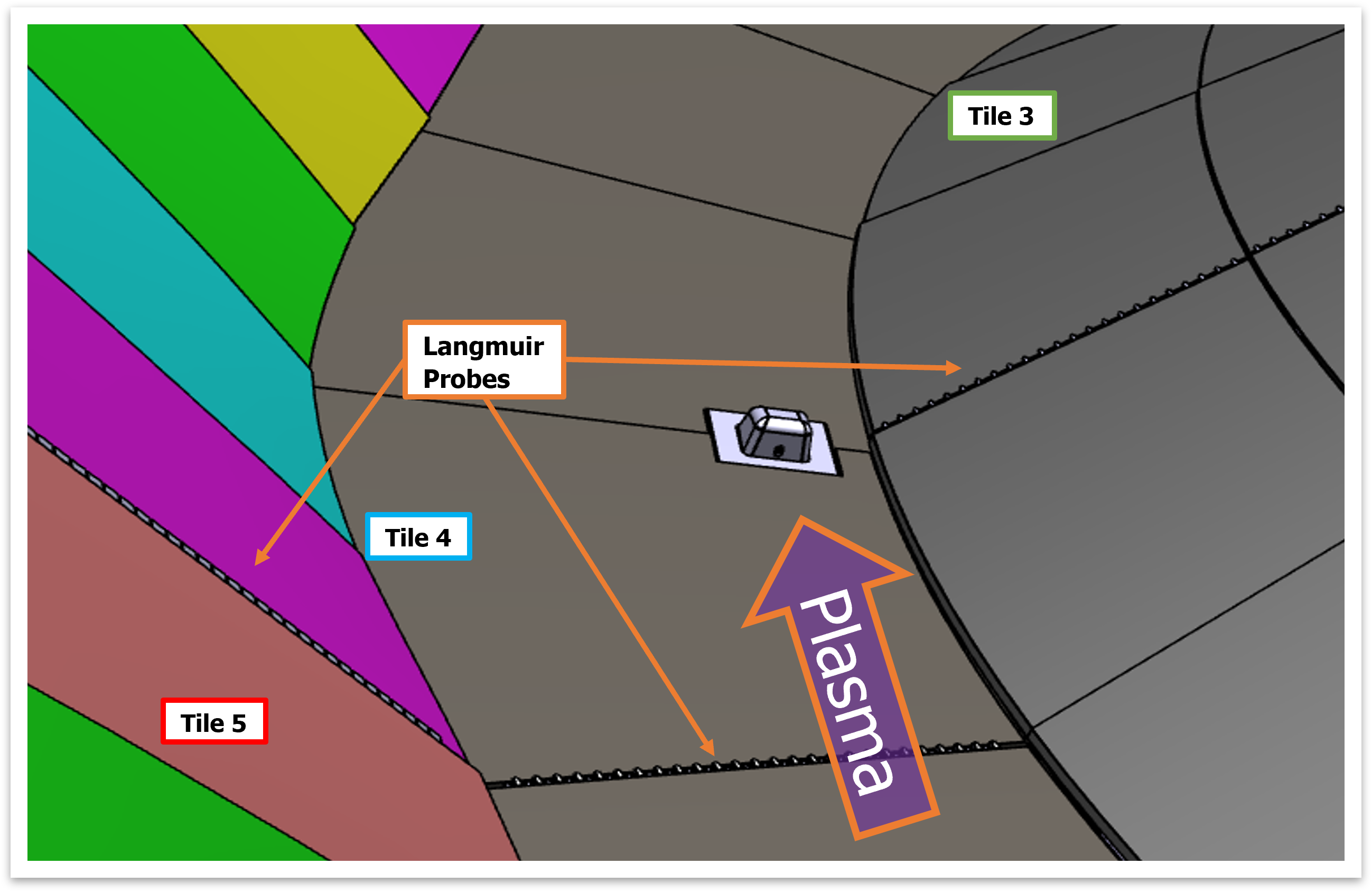}
\caption{CAD image showing the location of the DSF in the MAST-U divertor, with the RFEA mounted in place.}
\label{fig:DSF}
\end{figure}

This study focuses on the novel application of a Retarding Field Energy Analyzer (RFEA) integrated within MAST-U’s divertor system. The RFEA is mounted in the Divertor Science Facility (DSF). The DSF is an experimental platform located in the Tile 4 of MAST-U divertor and allows different probe heads, including the RFEA, to be inserted into the divertor region (see \autoref{fig:DSF}). This setup enables measurements as the plasma strike point sweeps over the DSF location.
The RFEA was employed to perform ion temperature ($T_{i}$) measurements across different plasma scenarios, including the transition from Conventional Divertor (CD) to Super-X Divertor (SXD)\cite{osborne2023initial}. 
Fluctuations in $T_{i}$ significantly impact plasma-facing components (PFCs) \cite{riccardi2011preliminary}. While $T_{i}$ fluctuations reflect changes in ion energy, heat flux fluctuations represent the overall power affecting PFCs. Managing these fluctuations is crucial, as exceeding the heat flux limit of 10 MW/m$^{2}$ can accelerate erosion and cause structural damage to materials. Understanding $T_{i}$ in the divertor region is therefore essential for developing protective strategies for PFCs, particularly because sputtering yield depends on the energy of incoming ions. These insights are vital for ensuring the longevity and performance of materials exposed to high-energy ion fluxes \cite{stangeby2000plasma}.

To fully assess these fluctuations, it is essential to measure $T_i$ not only during steady-state plasma conditions but also during transient events such as Edge Localized Modes (ELMs). ELMs are characterized by rapid bursts of heat and particle fluxes that enhance temperature fluctuations and can impact the material integrity of PFCs. Therefore, understanding the dynamics of $T_i$ during both ELM and inter-ELM phases can be helpful for developing effective ELM control strategies, as it provides insights into critical periods of heat flux, energy deposition, and the effectiveness of mitigation techniques. 
Elevated $T_{i}$ during inter-ELM phases influences processes such as particle recycling, impurity influx, and power exhaust, all of which impact the speed of pedestal recovery\cite{cavedon2019ion}.
The remainder of this paper is organized as follows: Section 2 provides an overview of the experimental setup and diagnostic integration, highlighting the RFEA and LP systems. Section 3 presents the results, focusing on \(T_i\) measurements across different plasma scenarios. Finally, Section 4 discusses the conclusions and future directions.

\section{Experimental Setup}
\begin{figure}[H]
% \centering
\includegraphics[width=0.48\textwidth]{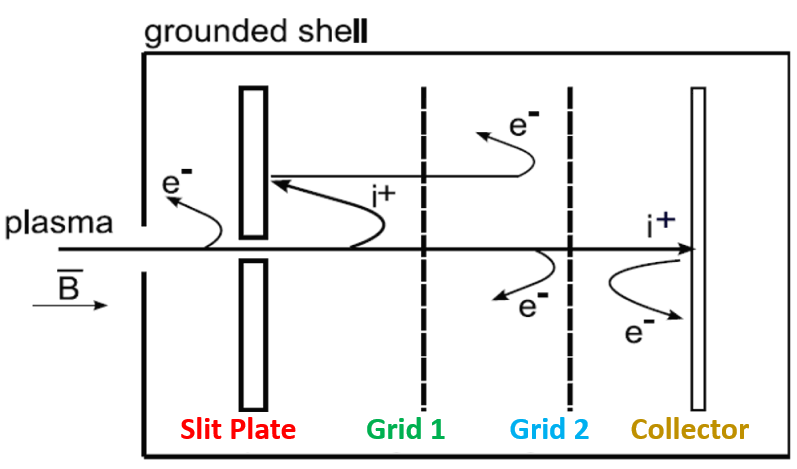}
\caption{Schematic of of the MAST-U RFEA module showing the function of the slit plate, grids
and collector plate\cite{elmore2012divertor}.}
\label{fig:RFEA_Grids}
\vspace*{-1mm}
\end{figure}

\begin{figure}[H]
\centering
\includegraphics[width=0.50\textwidth]{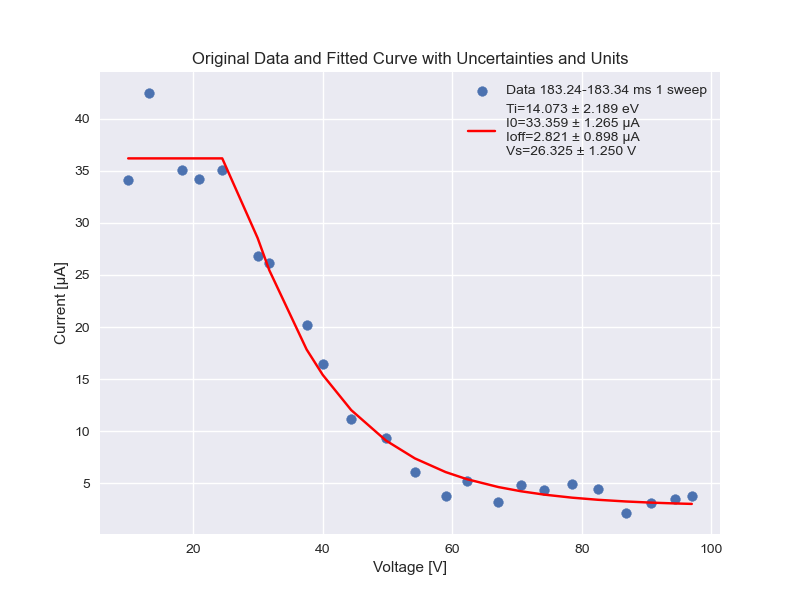}
\caption{Raw data points (blue) of ion current versus applied voltage, gathered from a single sweep between 183.24 to 183.34 ms during shot 49218. These data points have been fitted (red curve) to a theoretical model (Eq:\ref{Expanded_Fit_Equation}).}
\label{fig:Fit_example}
\end{figure}

The RFEA is a diagnostic specifically designed to capture and analyze the energy distribution of charged particles, predominantly ions, within a plasma. It has been used to measure ion temperature within the scrape-off layer (SOL) on tokamaks including MAST, ASDEX-Upgrade, Tore-Supra and JET \cite{kovcan2008reliability}\cite{pitts2005far}\cite{allan2016ion}. This device operates on the principle of electrostatic retardation of ions using a multi-grid system (\autoref{fig:RFEA_Grids}). The MAST-U RFEA consists of an entrance grid (Slit Plate), intermediate grids (Grid 1 and Grid 2), and a Collector Plate.
The entrance grid repels electrons, ensuring that primarily ions enter the device. The subsequent grids are biased at different potentials to create an electric field that selectively allows ions of specific kinetic energies to pass through. Grid 1 is a discriminating grid that sweeps from 0 to a positive voltage ($>$ 3$T_{e}$). Grid 2 is a suppression grid held at negative voltage to repel secondary electrons released by ions hitting the collector plate.
Grid 2 is a suppression grid held at negative voltage to repel secondary electrons released by ions hitting the collector plate. It also suppresses fast electrons that manage to pass through the slit plate, as well as secondary electrons originating from the slit plate itself \cite{kovcan2008reliability}.
Data collection with the RFEA involves capturing the ion current at the 
collector plate as a function of the retarding potential applied to 
Grid 1. This process generates an I-V characteristic curve, from which 
the ion temperature is derived using \autoref{Expanded_Fit_Equation}, where $I_{0}$ is the ions saturation current, $V_{s}$ the sheath voltage and $I_{off}$ accounts for any current offsets due to the electronics of the system. \autoref{fig:Fit_example} shown an I-V characteristic of a plasma as measured by the RFEA.

\begin{equation}
\label{Expanded_Fit_Equation}
    I_{\text{col}} = \begin{cases} 
    I_0 + I_{\text{off}}, & \text{if } V_{\text{grid1}} \leq V_s \\
    I_0\exp\left(-\frac{V_{\text{grid1}} - V_s}{T_i}\right) + I_{\text{off}}, & \text{if } V_{\text{grid1}} > V_s 
    \end{cases}
\end{equation}

\begin{figure*}[ht]
\centering
\includegraphics[width=1\textwidth]{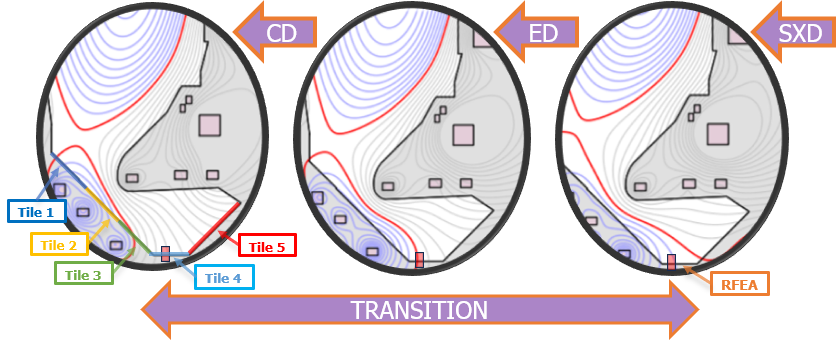}

\caption[Transition from CD to ED and SXD]{Transition from Conventional Divertor (CD) to Elongated Divertor (ED) and Super-X Divertor (SXD) based on EFIT reconstructions. Divertor tiles (Tile 1–5) and the RFEA location are highlighted.}
\label{Transition}
\end{figure*}

The primary data recorded includes ion current magnitudes at varying retarding potentials, which are processed using a Python-based analysis tool. This tool integrates the raw RFEA data and applies a fitting algorithm to deduce the ion temperature and density. To reconstruct  the $T_i$ profiles around the strike point, the dynamic motion of the strike point in the MAST-U divertor chamber is exploited.
In this work, two configurations were examined that allow measurable signals on the RFEA: the Elongated Divertor (ED), and transition from CD to SXD (\autoref{Transition}).
The ED configuration extends the outer leg of the plasma from a CD to a chosen position on Tile 4.
When the strike point is positioned at the RFEA location, the temporal evolution of $T_{i}$ can be measured. This allows for the measurement of multiple ELM temperature profiles in H-mode\cite{elmore2016scrape}. 
Another approach involves transitioning from the CD to the SXD scenario, during which the radial profile is measured as the strike point sweeps over the RFEA. As the strike point passes over the RFEA, this provides a window of approximately 20-30 ms for obtaining a measurable signal. Temporal profiles recorded during this sweep can then be converted into radial distances from the Last Closed Flux Surface (LCFS) to the target (probe position), denoted as $\Delta R^{tgt}_{LCFS}$.

\begin{table*}[ht]
\centering
\caption{Summary of  Shots}
\label{tab:Shots1}
\begin{tabular}{|c|c|c|c|c|c|}
\hline
\textbf{Shot} & \textbf{$I_p$} [kA] & \textbf{$P_{\text{NBI}}^{tot}$} [MW] & \textbf{time} [ms] & \textbf{$n_e^{\text{core}}$ } [$m^{-3}$] & \textbf{ Configuration} \\ \hline
49218         & 750              & 3.1  & 170 - 204  & $(2.2 - 2.92) \times 10^{19}$ & CD to SXD \\ \hline
49138         & 750              & 3.2  & 609 - 698          & $(4.1 - 4.4) \times 10^{19}$ & ED\\ \hline
\end{tabular}
\end{table*}

In the experiments conducted, the plasma current was maintained around 750 kA, with core electron densities ranging from $2.2 - 2.92 \times 10^{19} \, \text{m}^{-3}$ for shot 49218 and $4.1 - 4.4 \times 10^{19} \, \text{m}^{-3}$ for shot 49138. The Neutral Beam Injection (NBI) power varied between 3.1 to 3.2 MW across both shots. Further details of the experimental shots are summarized in \autoref{tab:Shots1}. Shot 49218 features $T_i$ measurements during the transition from CD to SXD in H-mode, while shot 49138 captures multiple ELM events within the ED configuration. 

\section{Results}
This study presents a comparative analysis of ion saturation current density ($J_{sat}$) and temperatures across the SOL and PFR for shot 49218, utilizing the RFEA for ion measurements and the LP for electron temperature ($T_e$) data \cite{ryan2023overview}. Ion temperature ($T_i$) profiles are compared with $T_e$ measurements at the same radial positions during the CD to SXD transition. Additionally, the study discusses the ion temperature measurements of ELM phenomena recorded in shot 49138 in ED configuration.

\subsection{Conventional to Super X Divertor}
\begin{figure}[H]
\centering
\includegraphics[width=0.48\textwidth]{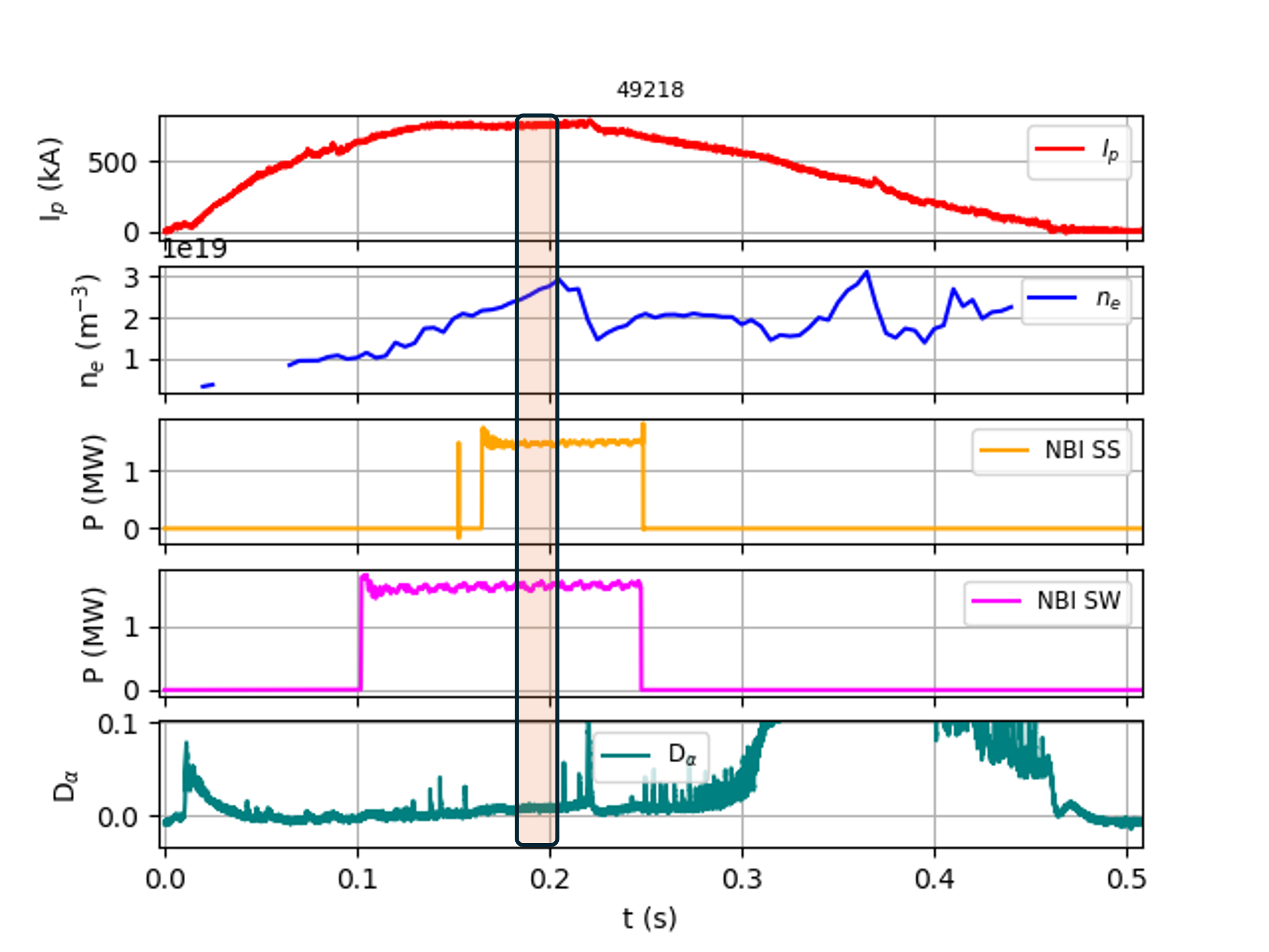}
\caption[Time trace of shot 49218]{Time trace of shot 49218, highlighting the period of measurement by the RFEA (shaded area). The plot shows key parameters including plasma current (\(I_p\)), line average core density (\(n_e\)), NBI power (\(P_{\text{NBI}}\)) and $D_{\alpha}$ signal from the midplane.}
    \label{fig:shot_49218_time_trace}
\end{figure}

\begin{figure}[H]
\centering
\includegraphics[width=0.48\textwidth]{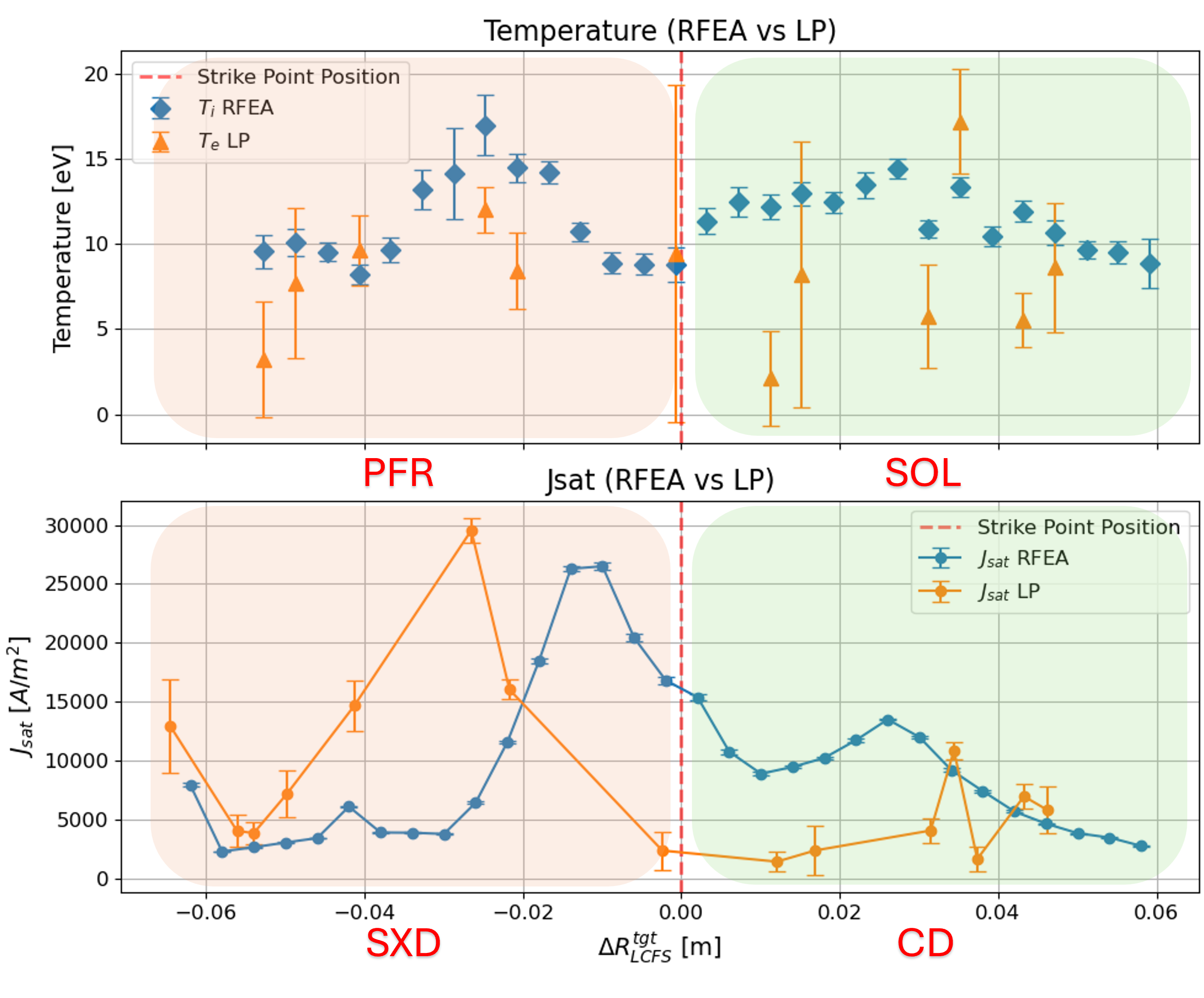}
\caption{Comparative analysis of $T_{i}$, $T_{e}$, and $J_{sat}$ as a function of distance to the LCFS for shot 49218. The top panel shows the radial distribution of $T_{i}$ (diamonds) and $T_{e}$ (triangles) with propagated error bars. The bottom panel compares $J_{sat}$ from RFEA and LP. Measurements in the PFR (distances $<$ -0.04 m) may be unreliable due to transient phenomena. Divertor configuration changes are annotated: 'CD' for Conventional Divertor (SOL) and 'SXD' for Super-X Divertor (PFR). Data range: 170-204 ms.}
\label{fig:Comparison_JSAT_T}
\end{figure}

\begin{figure}[H]
\centering
\includegraphics[width=0.40\textwidth]{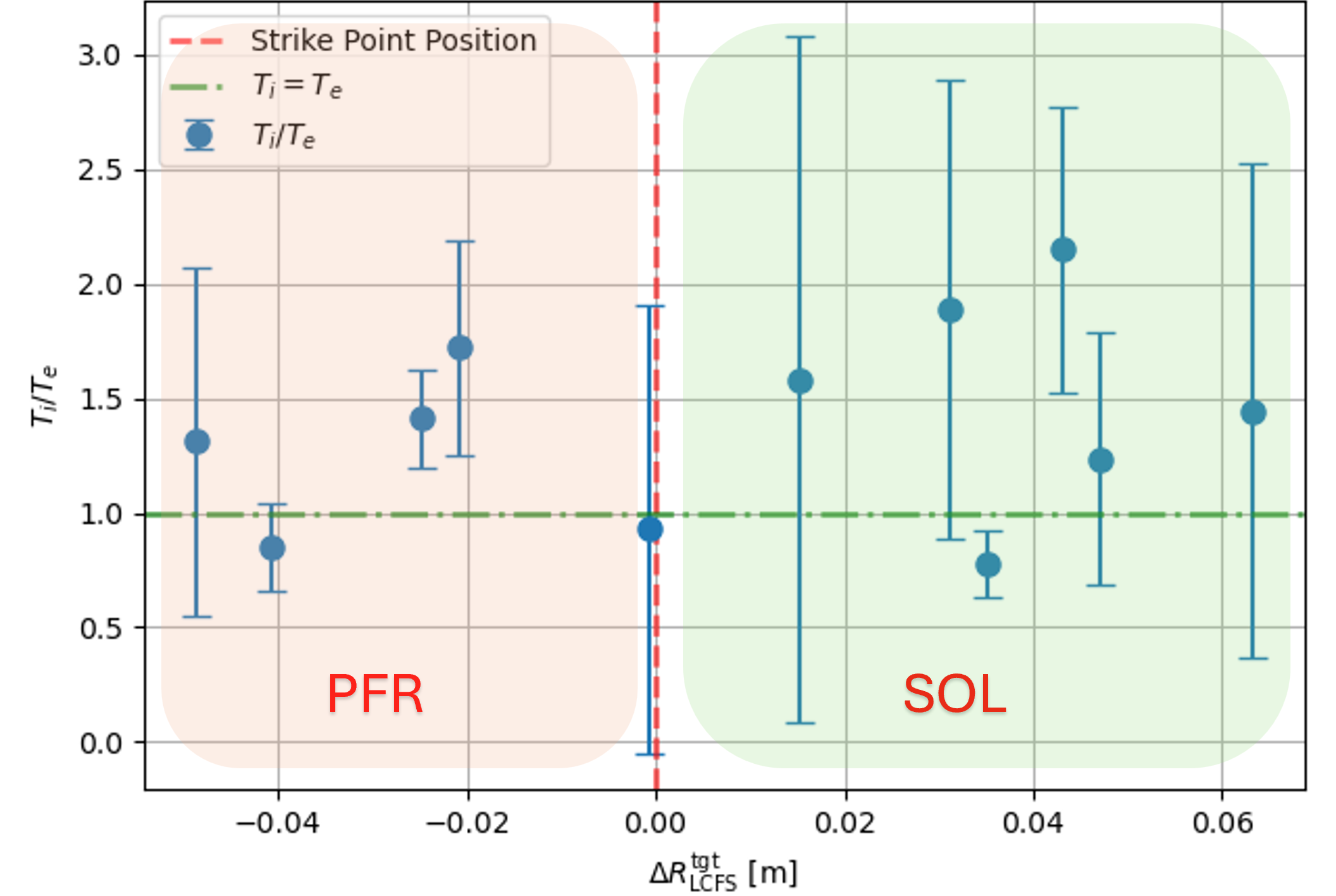}
\caption{Variation of the ion-to-electron temperature ratio ($T_{i}$/$T_{e}$) measured by the RFEA and  LP, as a function of radial distance during a strike point sweep.}
\label{fig:Ti_Te_Ratio}
\end{figure}

For shot 49218, the time trace is shown in \autoref{fig:shot_49218_time_trace}, providing an overview of the temporal evolution. \autoref{fig:Comparison_JSAT_T} presents temperature and current profiles, where the upper subplot shows the ion temperature ($T_{i}$) measured by the RFEA alongside the electron temperature ($T_{e}$) obtained from the LP, and the lower subplot compares the ion saturation current ($J_{sat}$) from both diagnostics.
RFEA data is radially binned to match the LP time resolution. The binned data are calculated as the mean of the data points within each bin. The error for each bin is calculated using individual errors of the data points within the bin. This approach ensures that both the bin's data and their associated errors are accurately represented. The $T_{i}$ profile correlates well with the $J_{sat}$ from the RFEA, with peak $T_{i}$ around 17 eV in the PFR.
The peaks in $T_{i}$ and $J_{sat}$ are not necessarily expected to coincide. Ion temperature ($T_{i}$) reflects the thermal energy of ions, while ion saturation current ($J_{sat}$) is related to the ion flux reaching the probe. 
As a result, their peaks may be spatially offset, reflecting distinct behaviors in ion energy and particle flux.
Nevertheless, the general congruence between the $T_{i}$ profile shape and $J_{sat}$ is maintained, with both showing a double peak. In this type of measurement, performing a strike point scan without altering plasma conditions is not feasible, as the scan is conducted during the transition from the CD to SXD configuration. This complicates the interpretation of the profiles. 
During the transition, the plasma conditions in the divertor evolve, affecting both particle and heat fluxes. As a result, ion temperature profiles might be affected by changes in confinement, particle recycling, and plasma-material interactions. The double-peaked $J_{sat}$ profile likely results from evolving plasma conditions, where localized heating and transport processes change as the divertor geometry shifts. The transition between configurations can create distinct regions of increased ion energy and flux, leading to the observed double-peaked structure.
Comparatively, the $J_{sat}$ profile from the LP exhibits some similar features to that of the RFEA. Both profiles show a prominent peak in the PFR and a smaller secondary peak in the SOL. However, the LP profile shows a larger magnitude in the PFR compared to the RFEA, indicating a higher measured ion flux at these radial positions. This discrepancy could be partly due to uncertainty in the collection areas utilized by the diagnostics to calculate $J_{sat}$ and the influence of the toroidal field (TF) ripple \cite{arter2014cad}. The TF ripple introduces differences in the magnetic field magnitudes at the respective toroidal locations of the probes, leading to local variations in plasma parameters and, consequently, differences in ion saturation current profiles. Specifically, the TF ripple can cause perturbations in the magnetic field structure that affect ion confinement and transport. These perturbations may locally enhance or suppress ion fluxes, resulting in the large peak observed in the $J_{sat}$ profile measured by the LP in the PFR. In contrast, the RFEA, located at a different toroidal position, may experience a slightly altered magnetic field configuration, leading to a reduced ion flux compared to the LP.
The red dashed line on \autoref{fig:Comparison_JSAT_T} and \autoref{fig:Ti_Te_Ratio}, marks the strike point location on the RFEA orifice, serving as a key reference point to distinguish between the SOL (positive $\Delta R_{LCFS}^{tgt}$, to the right) and the PFR (negative $\Delta R_{LCFS}^{tgt}$, to the left).

During the strike point sweep from CD to SXD, the RFEA probe initially encounters the SOL as the strike point approaches, and subsequently measures the PFR as the strike point passes over, providing a sequential view of plasma conditions across these regions. In the PFR for distances less than -0.04 meters, transient phenomena are observed in both the $D_{\alpha}$ signal and the RFEA slit plate current. Consequently, the measurements in this region may not be reliable for this study.
As shown in \autoref{fig:Ti_Te_Ratio}, the highest measured $T_{i}/T_{e}$ ratio occurs at 2.2 in the SOL at approximately 0.04 m to the right of the strike point, with values tapering off towards the PFR and the far SOL.
Moreover, the larger error bars observed in these regions stem from the error bars associated with the uncertainty of the LP $T_{e}$ measurement, compared to the RFEA $T_{i}$. For this shot, the LPs were set in multiplexer mode\cite{ryan2023overview}, implying a lower time resolution. This exposes the LP probe to different plasma conditions within a single voltage sweep and does not allow to bin them to reduce the uncertainty, as is possible with the RFEA. 
% The highest measured $T_i/T_e$ ratio occurs in the SOL around 0.04 m from the strike point, with values gradually decreasing towards the PFR and the far SOL, as shown in \autoref{fig:Ti_Te_Ratio}. 

These findings underscore the influence of localized plasma conditions and magnetic field configurations on $T_{i}$ and $T_{e}$. The higher $T_{i}$ relative to $T_{e}$ in certain areas could be indicative of less efficient ion cooling processes in these regions, potentially due to decreased ion interactions with neutrals or reduced efficacy of radiative cooling mechanisms\cite{rosmej2011nonequilibrium}. These phenomena could be related to the level of detachment observed during these transient measurements, which affects both ion and electron temperatures.
Further research is necessary to elucidate 
the specific mechanisms contributing to the 
observed temperature disparities. Detailed 
modeling and simulation efforts should be 
directed towards understanding the impact of 
toroidal magnetic field ripple effects\cite{arter2014cad}.
Additionally, the experimental setups could be 
optimized to reduce measurement uncertainties 
and enhance the resolution of temperature 
profiles from LP (e.g. not multiplexing).

In this study, $T_{i}$/$T_{e}$ ratios ranging from 1 to 2.2 were observed, indicating significant differences in the way energy is deposited and distributed within the plasma, comparable to previous studies at the midplane carried out by Ko{\v{c}}an et al.\cite{kovcan2008edge}. Furthermore prior studies on MAST, such as those by Elmore et al.\cite{elmore2012upstream}, have reported similar ion temperature measurements up to 15 eV on the divertor, under conditions of lower plasma density and current (450 kA) in ohmic heated CD plasmas. This recent investigation on MAST-U thus provides additional evidence of the effectiveness of the cooling impact afforded by the novel extension of the outer strike point leg into a Super-X divertor configuration.

\subsection{ELMs Measurements in Elongated Divertor}
\begin{figure}[H]
\centering
\includegraphics[width=0.48\textwidth]{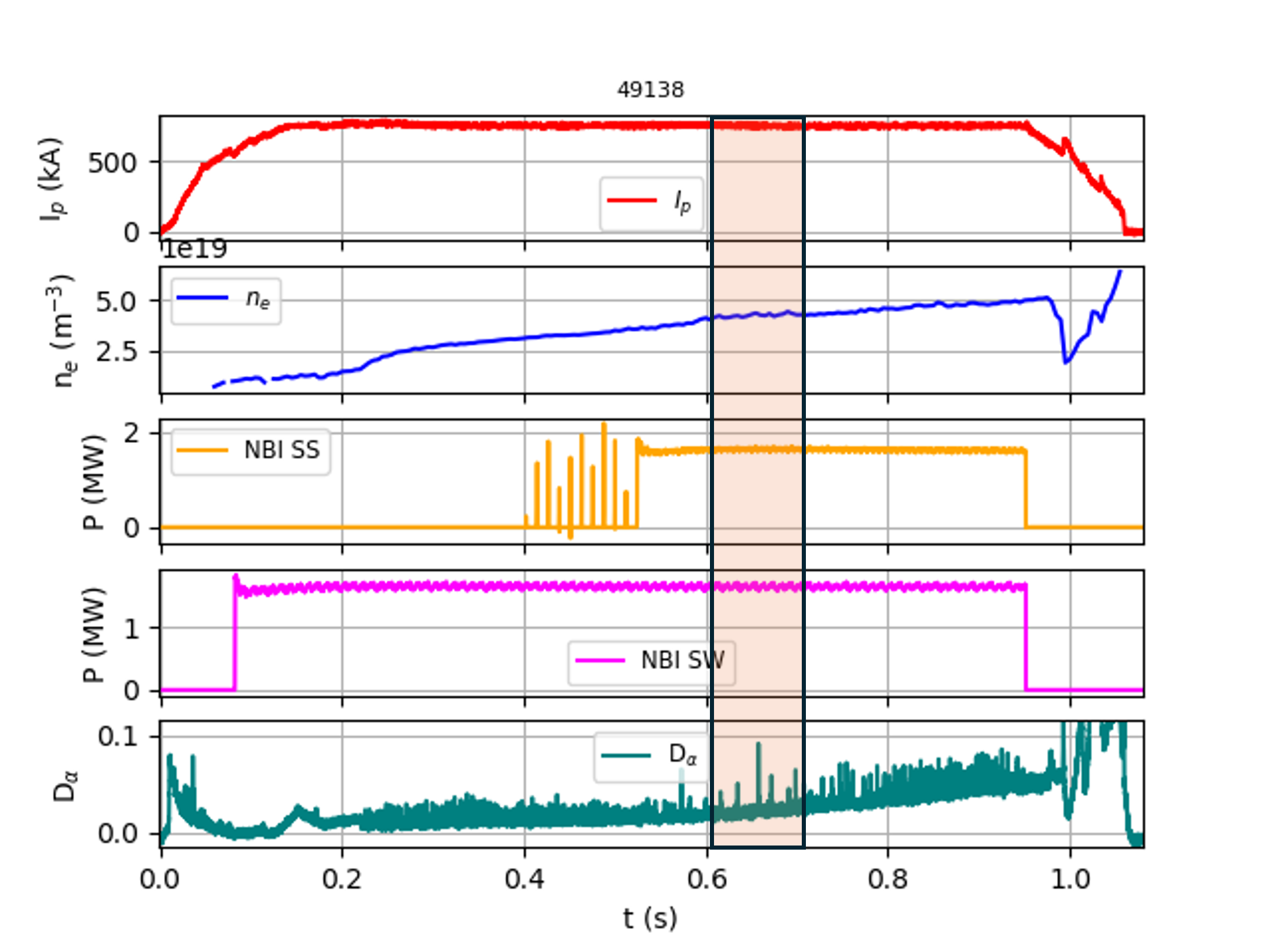}
\caption[Time trace of shot 49138]{Time trace of shot 49138, highlighting the period of measurement by the RFEA (shaded area). The plot shows key parameters including plasma current (\(I_p\)), line average core density (\(n_e\)), NBI power (\(P_{\text{NBI}}\)) and $D_{\alpha}$ signal from the midplane.}
    \label{fig:shot_49138_time_trace}
\end{figure}

\begin{figure}[H]
\centering
\includegraphics[width=0.40\textwidth]
{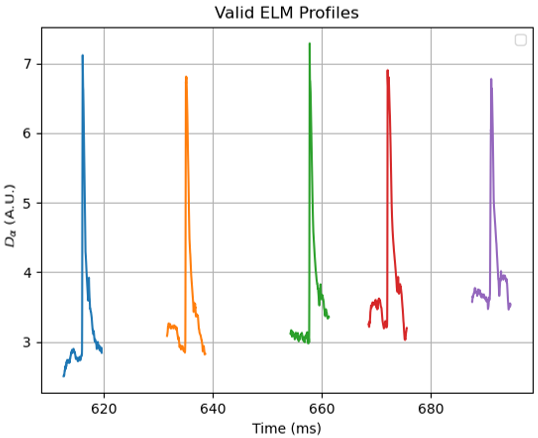}
\caption{ELM 
instances used in this experiment (each represented by a different 
color) captured over time. These profiles are 
used to compile and analyze the average 
characteristics of ELM behavior within the 
experiment.}
\label{Valid_ELM}
\end{figure}

\begin{figure}[H]
\centering
\includegraphics[width=0.45\textwidth]{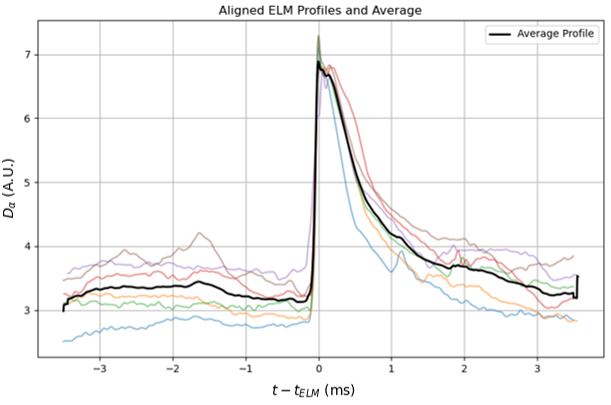}
\caption{ELM instances from \autoref{Valid_ELM} (colored lines) aligned with respect to a common onset time ($t - t_{\text{ELM}} = 0$), overlaid with their collective average profile (thick black line). The alignment highlights the temporal consistency and variability of ELM characteristics across different events.}
\label{Aligned_ELM}
\end{figure}

The fast sweep method employed by the RFEA 
enhances the temporal resolution of 
measurements to 100 $\mu s$. A similar approach has been previously used in ASDEX Upgrade\cite{ochoukov2020ion}. This 
capability is crucial for capturing the rapid 
dynamics of ELMs. By 
achieving such a high temporal resolution, 
some instance of an ELM can be recorded.
To reconstruct a comprehensive profile of 
these transient events, an ELM 
identification technique akin to those 
described in prior studies by Kirk et al.\cite{kirk2005structure} and Elmore et al.\cite{elmore2016scrape} was implemented. This method involves 
aligning multiple ELM instances on the same 
temporal scale and aggregating them to form a 
composite profile. The onset of an ELM is 
determined by the time at which 
the midplane $D_{\alpha}$ signal rises to 10\% of the peak value above the inter-ELM level. The time trace for shot 49138 is shown in \autoref{fig:shot_49138_time_trace}, providing an overview of key parameters during the measurement period. 
\autoref{Valid_ELM} displays selected ELM events during shot 49138, in the range from 609 to 698 ms, chosen to represent Type I ELMs. Some ELMs with a current exceeding 3A, which surpassed the power supply saturation limit, were excluded from the analysis.
Each color in \autoref{Aligned_ELM} represents a different ELM, illustrating the variability in their temporal characteristics and amplitude. By aligning these instances using a time scale $t - t_{\text{ELM}}$, where $t_{\text{ELM}}$ denotes the ELM onset just identified, an averaged ELM profile was constructed. This approach ensures that each ELM, occurring at different times during the experiment, is synchronized to a common relative time, to create the coherent averaged profile in black in 
\autoref{Aligned_ELM}.
% , shows the 
% averaged profile, facilitating a synchronized 
% view of the ELM shape.

The aligned ELM measurements were taken within a range of $\pm 3$  ms around the ELM peak. 
During the measurements, the strike point was consistently maintained at the RFEA position, allowing for the measurement of the temporal evolution of ELMs at a fixed spatial location.
This setup ensures that the measured temperature profiles reflect the dynamics at the point of impact, providing a reliable basis for understanding the thermal stresses imposed by ELMs on the divertor materials.

\begin{figure}[H]
\centering
\includegraphics[width=0.49\textwidth]{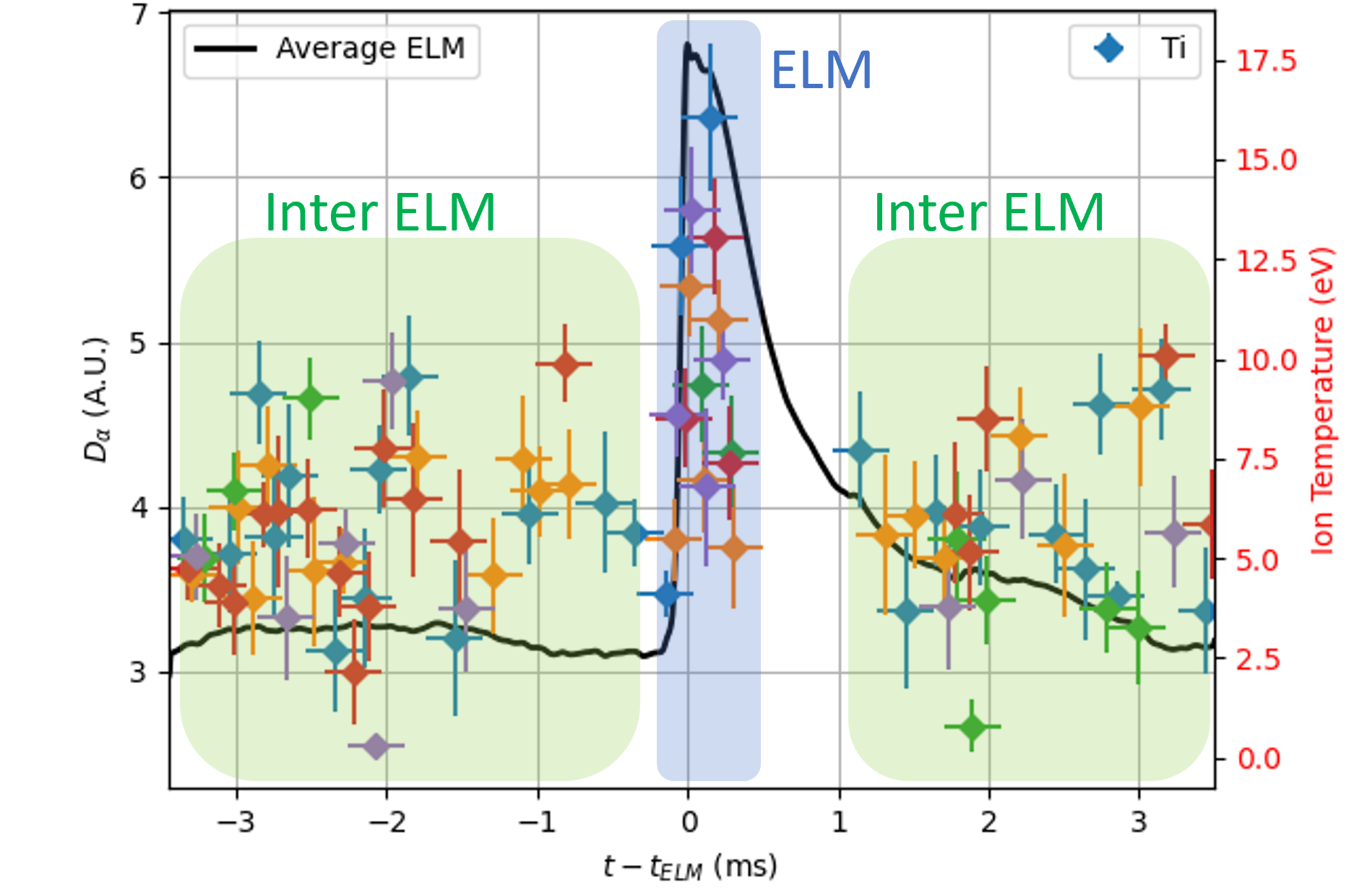}
\caption[Average ELM profile]{Average ELM profile (black line) along with individual temperature measurements during peak ELM events (blue area) and inter-ELM periods (green area) in the time range 609-698 ms.}
\label{ELM_ti}
\end{figure}

\autoref{ELM_ti} illustrates the temporal 
variation in $T_{i}$ during the ELM and inter-ELM periods, captured 
by the RFEA.
During the ELM, the $T_{i}$ reaches a peak of $16.03 \pm 1.84 \, 
\text{eV}$, almost 3 times higher than the inter ELM average of $5.73 
\pm 0.15 \, \text{eV}$. This elevated temperature correlates with the higher particle
and energy fluxes to the divertor region, driven by the rapid release 
of energy during the ELM event. The peak area considered is marked by 
the blue rectangle in the plot, denoting the peak temperature 
measured from all the considered ELMs. This profile exhibit a rapid ascent in $T_{i}$ coinciding with the onset of the ELM, followed by a steep decline immediately after the ELM peak. The  gap from about 0.4 to 1 ms is due to the lack of collector current signal, which collapses to near zero after each ELM peak. The signal level returns to analyzable values at approximately +1 ms. This behaviour could be consistent with the space charge limitation previously found in MAST RFEA data \cite{elmore2013scrape}. 
The space charge limitation, caused by the build-up of charge in the analyzer head, partially offsets the applied electric field, impacting measurement accuracy. While lowering the slit plate voltage may help reduce these effects, it is not sufficient for complete mitigation. Optimizing the RFEA settings, such as increasing power supply current capacity and refining slit plate bias remains necessary to effectively minimize space charge limitations and ensure reliable data collection.
An alternative explanation is that the heat and particle fluxes just after the ELM crash are very low, as most of the available heating is being used to rebuild the pedestal. This could explain why the collector current does not register during this interval. The signal only reappears once the particle fluxes increase again and reach the collector, allowing measurable current to be detected.
In the inter-ELM region, the $T_{i}$ stabilizes to a lower average of $5.73 \pm 0.15 \, \text{eV}$, as indicated by the green area ($-3.5 < t-t_{ELM} < -0.3$ and $1 < t-t_{ELM} < 3.5$). This lower 
temperature phase reflects a more quiescent state of the plasma, where 
energy deposition to the target is reduced, and the plasma conditions are 
relatively stable.

% The difference in the ion temperature between the two shots can be explained by the variation in plasma density. In shot 49138, the higher density results in increased collisionality, which leads to more efficient energy exchange and lower ion temperature. In contrast, in shot 49218, the lower density results in fewer collisions, allowing the ions to retain more energy and resulting in a higher $T_i$. This behavior highlights the critical role that plasma density plays in determining energy distribution and ion temperature in the divertor region.

\section{Conclusions}

This study presents the first $T_{i}$ measurements obtained using the RFEA diagnostic in the MAST-U divertor region during two distinct shots: one in the CD configuration and another during the transition to the SXD configuration. During the CD to SXD transition, the $T_{i}$ radial profile exhibited two peaks on either side of the strike point, reaching 17 eV, before decreasing further away. These measurements were analyzed alongside $T_{e}$ data derived from LP at the same radial position. The comparison of $T_{i}$ and $T_{e}$ revealed non-equilibrium characteristics in the plasma, with $T_{i}$/$T_{e}$ ratios ranging from 1 to 2.2, consistent with previous midplane studies \cite{kovcan2008edge}. The observed variations in the $T_{i}/T_{e}$ ratio suggest complex interactions between ion and electron populations, highlighting areas for further investigation, particularly concerning energy transfer mechanisms in the plasma SOL. These mechanisms are crucial as they impact overall plasma behavior, including confinement quality and power exhaust efficiency.
The ED configuration enabled measurements of both peak and inter-ELM ion temperatures, with ELM peaks reaching 16.03 eV increasing to approximately three times above the inter-ELM average.
In summary, the implementation of the RFEA has enhanced the diagnostic capabilities of MAST-U, offering new insights into ion dynamics within the divertor region. Future research should extend these measurements to higher power and longer pulse scenarios, which are crucial for achieving the operational conditions expected in commercial fusion reactors.
Future $T_{i}$ measurements during ELMs at locations beyond the strike point, such as in the SOL and PFR, will help determine if the maximum $T_{i}$ occurs away from the strike point. These measurements could provide valuable insights into energy transport and the spatial distribution of ion temperature at the target during ELM events. Additionally, optimizing the RFEA settings in future experiments such as, increasing power supply capacity and refining slit plate bias settings, could improve the capture of the entire ELM profile. Space charge limitations were identified as a possible factor impacting ELM $T_{i}$ measurement accuracy, necessitating further optimizations. These efforts, coupled with refined plasma modeling based on these empirical data, would enhance predictive capabilities.

\section*{Acknowledgements}
\fontsize{8pt}{8pt}\selectfont
This work has been funded by the EPSRC Energy Programme, grant EP/S022430/1 and the University of Liverpool. 
This work is supported by US Department of Energy, Office of Fusion Energy Sciences under the Spherical Tokamak program, contract DE-AC05-00OR22725. 
For the purpose of open access, the author(s) has applied a Creative Commons Attribution (CC BY) licence (where permitted by UKRI, ‘Open Government Licence’ or ‘Creative Commons Attribution No-derivatives (CC BY-ND) licence’ may be stated instead) to any Author Accepted Manuscript version arising.

%\end{thebibliography}

% Now we need a bibliography:

% The rest of your document follows, formatted as two-column text...
\end{multicols}
% Your document ends here!

\begin{thebibliography}{7}

\bibitem{osborne2023initial}
N.~Osborne, K.~Verhaegh, M.~Bowden, T.~Wijkamp, N.~Lonigro, P.~Ryan,
  E.~Pawelec, B.~Lipschultz, V.~Soukhanovskii, T.~van~den Biggelaar, {\em
  et~al.}, ``Initial fulcher band observations from high resolution
  spectroscopy in the mast-u divertor,'' {\em Plasma Physics and Controlled
  Fusion}, vol.~66, no.~2, p.~025008, 2023.

\bibitem{riccardi2011preliminary}
B.~Riccardi, R.~Giniatulin, N.~Klimov, V.~Koidan, A.~Loarte, ``Preliminary results of the experimental study of PFCs exposure to ELMs-like transient loads followed by high heat flux thermal fatigue,'' {\em Fusion Engineering and Design}, vol.~86, no.~9-11, pp.~1665--1668, 2011.

\bibitem{cavedon2019ion}
M.~Cavedon, R.~Dux, T.~P{\"u}tterich, E.~Viezzer, E.~Wolfrum, M.~Dunne, E.~Fable, R.~Fischer, G.~F.~Harrer, F.~M.~Laggner, {\em et~al.}, ``On the ion and electron temperature recovery after the ELM-crash at ASDEX upgrade,'' {\em Nuclear Materials and Energy}, vol.~18, pp.~275--280, 2019.


\bibitem{stangeby2000plasma}
P.~C. Stangeby {\em et~al.}, {\em The plasma boundary of magnetic fusion
  devices}, vol.~224.
\newblock Institute of Physics Pub. Philadelphia, Pennsylvania, 2000.

\bibitem{kovcan2008reliability}
M.~Ko{\v{c}}an, J.~Gunn, M.~Komm, J.-Y. Pascal, E.~Gauthier, and G.~Bonhomme,
  ``On the reliability of scrape-off layer ion temperature measurements by
  retarding field analyzers,'' {\em Review of scientific instruments}, vol.~79,
  no.~7, p.~073502, 2008.

\bibitem{pitts2005far}
R.~Pitts, W.~Fundamenski, S.~Erents, Y.~Andrew, A.~Loarte, C.~Silva, J.-E.
  contributors, {\em et~al.}, ``Far sol elm ion energies in jet,'' {\em Nuclear
  fusion}, vol.~46, no.~1, p.~82, 2005.

\bibitem{allan2016ion}
S.~Allan, S.~Elmore, G.~Fishpool, B.~Dudson, M.~Team, E.~M. Team, {\em et~al.},
  ``Ion temperature measurements of l-mode filaments in mast by retarding field
  energy analyser,'' {\em Plasma Physics and Controlled Fusion}, vol.~58,
  no.~4, p.~045014, 2016.

\bibitem{elmore2012divertor}
S.~Elmore, J.~W. Bradley, A.~Kirk, S.~Allan, A.~Thornton, J.~Harrison, and
  P.~Tamain, ``Divertor ion temperature measurements on mast by retarding field
  energy analyser,'' {\em Bulletin of the American Physical Society}, vol.~57,
  2012.

\bibitem{elmore2016scrape}
S.~Elmore, S.~Allan, G.~Fishpool, A.~Kirk, A.~Thornton, N.~Walkden,
  J.~Harrison, M.~Team, {\em et~al.}, ``Scrape-off layer ion temperature
  measurements at the divertor target during type iii and type i elms in mast
  measured by rfea,'' {\em Plasma Physics and Controlled Fusion}, vol.~58,
  no.~6, p.~065002, 2016.

\bibitem{ryan2023overview}
P.~J. Ryan, S.~Elmore, J.~Harrison, J.~Lovell, and R.~Stephen, ``Overview of
  the langmuir probe system on the mega ampere spherical tokamak (mast)
  upgrade,'' {\em Review of Scientific Instruments}, vol.~94, no.~10, 2023.
\bibitem{rosmej2011nonequilibrium}
F.~B. Rosmej, V.~S. Lisitsa, ``Nonequilibrium radiative properties in fluctuating plasmas,'' {\em Plasma Physics Reports}, vol.~37, no.~6, pp.~521--530, 2011, doi:10.1134/S1063780X1106006X.


\bibitem{arter2014cad}
W.~Arter, V.~Riccardo, and G.~Fishpool, ``A cad-based tool for calculating
  power deposition on tokamak plasma-facing components,'' {\em IEEE
  Transactions on Plasma Science}, vol.~42, no.~7, pp.~1932--1942, 2014.

\bibitem{kovcan2008edge}
M.~Ko{\v{c}}an, J.~Gunn, J.~Pascal, G.~Bonhomme, C.~Fenzi, E.~Gauthier, and
  J.~Segui, ``Edge ion-to-electron temperature ratio in the tore supra
  tokamak,'' {\em Plasma Physics and Controlled Fusion}, vol.~50, no.~12,
  p.~125009, 2008.

\bibitem{elmore2012upstream}
S.~Elmore, S.~Allan, A.~Kirk, G.~Fishpool, J.~Harrison, P.~Tamain,
  M.~Ko{\v{c}}an, R.~Gaffka, R.~Stephen, J.~Bradley, {\em et~al.}, ``Upstream
  and divertor ion temperature measurements on mast by retarding field energy
  analyser,'' {\em Plasma Physics and Controlled Fusion}, vol.~54, no.~6,
  p.~065001, 2012.

\bibitem{ochoukov2020ion}
R.~Ochoukov, M.~Dreval, V.~Bobkov, H.~Faugel, A.~Herrmann, L.~Kammerloher,
  P.~Leitenstern, A.~U. Team, E.~M. Team, {\em et~al.}, ``Ion temperature
  measurement techniques using fast sweeping retarding field analyzer (rfa) in
  strongly intermittent asdex upgrade tokamak plasmas,'' {\em Review of
  Scientific Instruments}, vol.~91, no.~6, 2020.

\bibitem{kirk2005structure}
A.~Kirk, H.~Wilson, R.~Akers, N.~Conway, G.~Counsell, S.~Cowley, J.~Dowling,
  B.~Dudson, A.~Field, F.~Lott, {\em et~al.}, ``Structure of elms in mast and
  the implications for energy deposition,'' {\em Plasma physics and controlled
  fusion}, vol.~47, no.~2, p.~315, 2005.

\bibitem{elmore2013scrape}
S.~Elmore, S.~Allan, A.~Kirk, A.~Thornton, J.~Harrison, P.~Tamain,
  M.~Ko{\v{c}}an, J.~Bradley, M.~Team, {\em et~al.}, ``Scrape-off layer ion
  temperature measurements at the divertor target in mast by retarding field
  energy analyser,'' {\em Journal of Nuclear Materials}, vol.~438,
  pp.~S1212--S1215, 2013.


\end{thebibliography}
\end{document}